\documentclass[12pt,draftcls,onecolumn]{IEEEtran}
\usepackage[applemac]{inputenc}
\usepackage{epsfig}
\usepackage{epstopdf}
\usepackage{latexsym}
\usepackage{amsfonts}
\usepackage{amssymb}
\usepackage{amsmath}
\usepackage{amsbsy}
\usepackage{color,graphicx}
\usepackage{multirow}
\usepackage{epic}
\usepackage{algorithm}
\usepackage{algpseudocode,MnSymbol,ifsym}
\usepackage{amssymb,amsmath}
\usepackage{url}
\usepackage{dsfont}

\newtheorem{proposition}{\textbf{Proposition}}

\usepackage{caption}
\usepackage{subcaption}

\newcommand{\dv}{\mathbf} 
\newcommand{\mc}{\mathcal} 

\graphicspath{{./}{./fig/}}

\begin{document}

\title{Compute-and-Forward on a Multi-User Multi-Relay Channel}
\vspace{1cm}

\author{\vspace{0cm}
\authorblockN{ \small Mohieddine El Soussi \qquad Abdellatif Zaidi \qquad
Luc Vandendorpe\thanks{This work has been supported in part by the IAP BESTCOM project funded by BELSPO and FP7 project NEWCOM\#. }
\thanks{Mohieddine El Soussi and Luc Vandendorpe are with ICTEAM, Universit\'e catholique de Louvain, Place du Levant, 2, 1348 Louvain-la-Neuve, Belgium. Email: \{mohieddine.elsoussi,luc.vandendorpe\}@uclouvain.be}
\thanks{Abdellatif Zaidi is with Universit\'e Paris-Est Marne La Vall\'ee,
77454 Marne la Vall\'ee Cedex 2, France. Email:
abdellatif.zaidi@univ-mlv.fr}}}

\vspace{1cm}

\maketitle

\begin{abstract}

In this paper, we consider a system in which multiple users communicate with a destination with the help of multiple half-duplex relays. Based on the compute-and-forward scheme, each relay, instead of decoding the users' messages, decodes an integer-valued linear combination that relates the transmitted messages. Then, it forwards the linear combination towards the destination. Given these linear combinations, the destination may or may not recover the transmitted messages since the linear combinations are not always full rank. Therefore, we propose an algorithm where we optimize the precoding factor at the users such that the probability that the equations are full rank is increased and that the transmission rate is maximized. We show, through some numerical examples, the effectiveness of our algorithm and the advantage of performing precoding allocation at the users. Also, we show that this scheme can outperform standard relaying techniques in certain regimes.
\end{abstract}

\section{Introduction}\label{secI}

\IEEEPARstart{N}{etwork} coding is a promising technique for modern communication networks. It was first introduced by Ahlswede \textit{et al.} in \cite{ACLY00} for wired networks. It allows each intermediate node to send out a function of the received packets from multiple sources \cite{FF56}. In general, the function does not need to be linear; however, most of the research on network coding has focused on linear codes since they have some noticeable features, in particular simplicity (e.g., see \cite{FS07}). 

For wireless networks, lattice codes attract great attention since they are based on linear structured codes. Recently, ``Compute-and-forward" (CoF) strategy, which is based on lattice codes, has been proposed \cite{NG11a}. This strategy implements network coding, where the receivers, instead of decoding the transmitted messages, decode finite-field linear combinations of transmitted messages. A receiver that is given a sufficient number of linear combinations recovers the transmitted messages by solving a system of independent linear equations that relate the transmitted messages. This strategy has been considered for different communication systems including the two-way relay channel \cite{KMT08}, the Gaussian network \cite{NG11a} and the multiple access relay channel (MARC) \cite{SZV14}.    

\begin{figure}[!ht]
  \begin{center}
  \includegraphics[width=7cm,height=5cm]{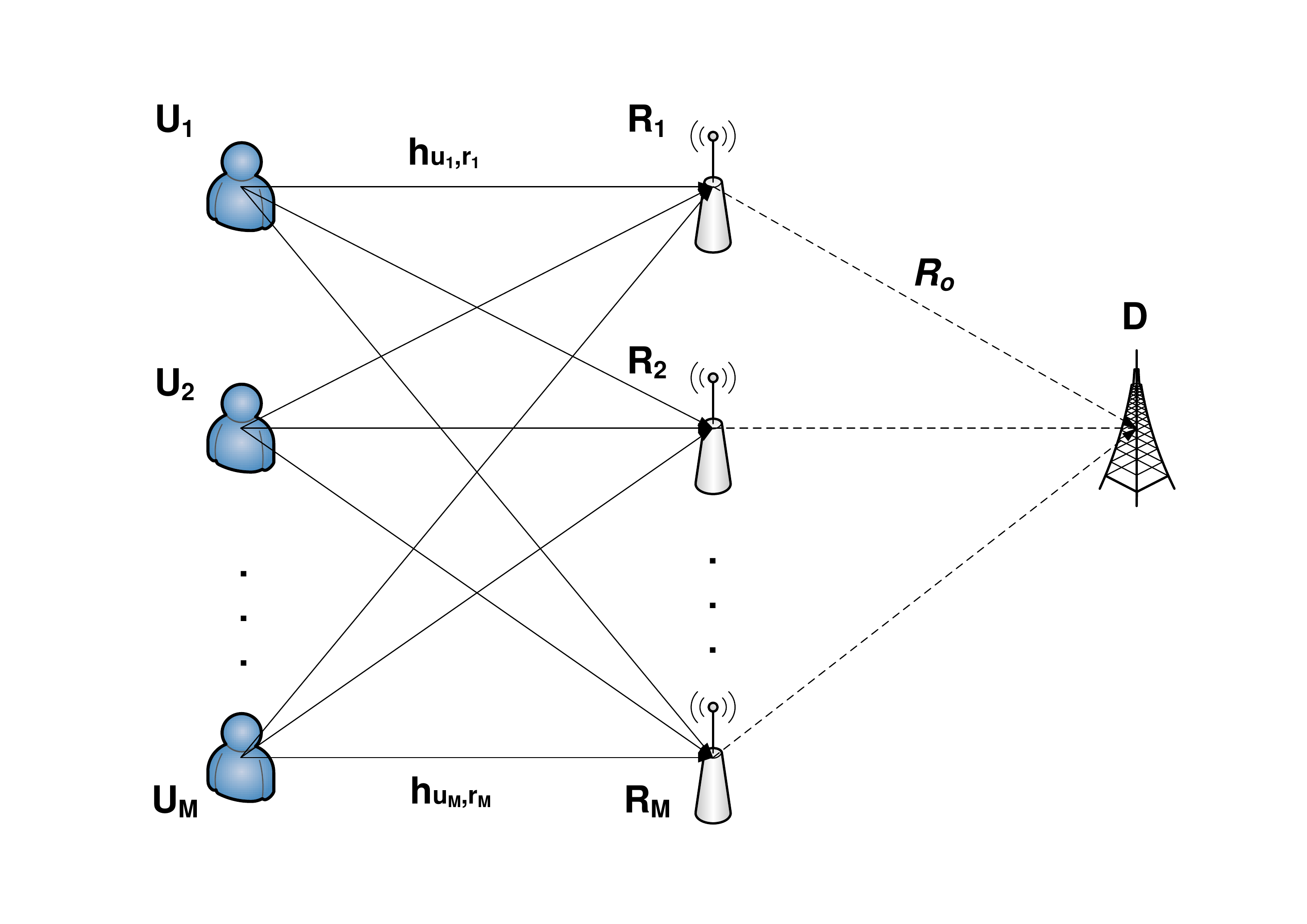}
  \end{center}
	\vspace{-0.75cm}
  \caption{Multi-user multi-relay network}
	\vspace{-0.3cm}
  \label{SystemModel}
\end{figure}

In this work, we consider a system where multiple users communicate with a destination with the help of multiple relays as shown in Figure~\ref{SystemModel}. The relays use CoF strategy where each relay decodes an integer-valued linear combination that relates the transmitted codewords and forwards it to the destination.


The multi-user multi-relay channel with CoF strategy has been considered in \cite{WC12, NG11a, CFB14}. In \cite{WC12}, the authors propose a method to compute the integer coefficient vectors of the linear combinations. In this method, the relays jointly optimize the integer vectors in such a way that the transmission rate is maximized and the matrix formed by these vectors is full rank. This method is not practical for large networks since additional signaling overhead is needed among the relays. In \cite{NG11a} and in contrast to \cite{WC12}, each relay independently (i.e. no coordination among the relays) computes an integer vector. Hence, there is a possibility that the received linear combinations at the destination are not full rank and thus the destination is not able to decode the transmitted messages. The authors in \cite{NG11a} propose a method that forces each relay to compute a linear combination with an integer coefficient that is different from zero. The integer coefficient that is different from zero varies from one relay to another. They showed that the probability of rank failure decreases at the expense of lower transmission rate. In \cite{CFB14}, the authors study the problem of maximizing the multicast throughput by properly allocating the resources (time and power) for given integer vectors. However, they assume that all users transmit with the same power.

In this work, we extend our previous work \cite{SZV14} to the case of multi-user and multi-relay. In \cite{SZV14}, we maximize the transmission rate by properly allocating the precoding factors (powers) at the users and computing the integer coefficients. The integer coefficients are jointly computed as in \cite{WC12}. In this setting, each relay independently computes a linear combination. We show that, by considering precoding allocation at the users, the probability of rank failure at the destination decreases since the precoding factor alters the channels between the users and the relays and hence alters the integer coefficient vectors. Thus, we aim to allocate the powers at the users in such a way that the received linear combinations at the destination are full rank and that the transmission rate is maximized.

\vspace{-0.75cm}

%
%

\section{System Model}\label{secI_subsecA}

We consider the communication system shown in Figure \ref{SystemModel} where a set of $M$ users $\text{U}_m$, $m = 1,\dots, M$, communicate with a destination with the help of $M$ relays. Each user $\text{U}_m$ wants to transmit a message $W_{u_m}$, which belongs to a set of alphabets $\mc W_{u_m}$, to the destination reliably in $2n$ uses of the channel. At the end of the transmission, the destination recovers the transmitted messages using its outputs. Let $R_{u_m}$ be the transmission rate of message $W_{u_m}$. We concentrate on the \textit{symmetric} rate case, i.e., $R_{u_m}=R_{\text{sym}} = R$. We divide the transmission time into two transmission periods with each of length $n$ channel uses. Also, we assume that the users are unable to communicate directly with the destination and the relays operate in a half-duplex mode.

%

During the first transmission period, each user $\text{U}_m$ encodes its message $W_{u_m} \in [1,2^{2nR}]$ into a codeword $\dv x_{u_m}$ using nested lattice codes \cite{NG11a} \cite{SZV14} and sends it over the channel. Let $\dv y_{r_m}$ be the signal received respectively at relay $m$ during this period. This signal is given by
\begin{eqnarray}
\mathrm{\mathbf{y}}_{r_m} &=& \sum_{i=1}^M{ h_{u_i,r_m}\mathrm{\mathbf{x}}_{u_i}} + \mathrm{\mathbf{z}}_{r_m}\nonumber
\label{outputs-relay-destination-first-transmission-period}
\end{eqnarray}
where \mbox{$h_{u_i,r_m}$} is the channel gain on the link between user $\text{U}_i$ and relay $m$, and $\mathrm{\mathbf{z}}_{r_m}$ is additive background noise at relay $m$.

During the second transmission period, each relay forwards the decoded linear combination to the destination through its own bit pipe with rate $R_o$ bits per channel use.

Throughout the paper, we assume that all channel gains are real-valued and fixed. We also assume that the users have full channel state informations (CSI) and that relay $m$ only knows the channel vector $\dv h_m=[h_{u_1,r_m},\: h_{u_2,r_m},\dots,\: h_{u_M,r_m}]^T \in \mathbb{R}^{M}$ to itself; and the noises at the relays are independent among each others, and independently and identically distributed (i.i.d) Gaussian, with zero mean and variance $N$. Furthermore, we consider the following individual constraints on the transmitted power (per codeword), 
\begin{align}
\mathbb{E}[\|\dv x_{u_m}\|^2]=n\beta^2_{u_m}P \leq nP_{u_m}, \qquad m=1,\dots,M
\end{align}
where $P_{u_m} \geq 0$ is a constraint imposed by the system; $P \geq 0$ is given, and $\beta_{u_m}$ is the precoding factor that can be chosen to adjust the actual transmitted power, and is such that $0 \leq |\beta_{u_m}| \leq \sqrt{P_{u_m}/P}$. 

\subsection{Notations}
The following notations are used throughout the paper. For convenience, we use the shorthand vector notation $\boldsymbol{\beta}=[\beta_{u_1},\: \beta_{u_2},\dots, \beta_{u_M}]^T$ $\in \mathbb{R}^{M}$. We also use $\boldsymbol{\beta} \circ \dv h_m \in \mathbb{R}^M$ to denote the Hadamard product of $\boldsymbol{\beta}$ and $\dv h_m$, $\dv I_n$ to denote $n$-by-$n$ identity matrix and $\text{rank}(\dv X)$ to denote the rank of matrix $\dv X$. Finally, we assume that logarithms are taken to base $2$; and, for $x \in \mathbb{R}$, $\log^{+}(x) := \max\{\log(x), 0\}$.

\section{Compute-and-Forward Strategy}\label{secI_subsecB}


The following proposition provides an achievable symmetric-rate for the multi-user multi-relay model that we study.
\vspace{0.1cm}
\begin{proposition}\label{proposition-achievable-sum-rate-compute-and-forward-at-relay}
For any set of channel matrix $\dv H = [\dv h_1,\:\dv h_1,\dots,\:\dv h_M]^T \in \mathbb{R}^{M\times M}$, the following symmetric-rate is achievable for the model that we study \cite[Theorem 5]{NG11a}\cite[Proposition 1]{SZV14}:
\begin{eqnarray}
R^{\text{CoF}}_{\text{sym}} = \: \max_{\{\dv a_m\}_{m=1}^M,\boldsymbol{\beta}}\:\: \min \left\{ \min_{m}\:R(\dv a_m,\dv h_m,\boldsymbol{\beta}),\: R_o \right\},
\label{achievable-sum-rate-compute-and-forward-at-relay}
\end{eqnarray}
where the maximization is over the precoding vector $\boldsymbol{\beta}$ such that $0 \leq |\beta_{u_m}| \leq \sqrt{P_{u_m}/P}$ and over the integer coefficients $\dv a_m \in \mathbb{Z}^M$, $m = 1,\dots,M$, such that $\text{rank}(\dv A)=M$, $\dv A=[\dv a_1,\: \dv a_2,\dots, \dv a_M]^T$ $\in \mathbb{Z}^{M \times M}$, and $R(\dv a_m,\dv h_m,\boldsymbol{\beta})$ is given by
\small{
\begin{equation}
R(\dv a_m,\dv h_m,\boldsymbol{\beta}) = \log^{+}\left(\left(\|\dv a_m\|^2-\frac{P((\boldsymbol{\beta} \circ \dv h_m)^T \dv a_m)^2}{N+P\|\boldsymbol{\beta} \circ \dv h_m\|^2}\right)^{-1}\right).
\end{equation}}
\end{proposition}
\normalsize
In this strategy, each relay independently computes a linear combination that relates the users' codewords and then forwards it to the destination. The destination recovers the transmitted messages only if the received linear combinations are full rank. In order to increase the probability that the received linear combinations are full rank and to maximize the transmission rate of Proposition~\ref{proposition-achievable-sum-rate-compute-and-forward-at-relay}, we develop at the users an iterative algorithm that finds the optimum precoding vector $\boldsymbol{\beta}$.

\section{Symmetric Rate Optimization}\label{secII}

The following section is devoted to finding optimal precoding and integer-coefficients that maximize the symmetric-rate of Proposition~\ref{proposition-achievable-sum-rate-compute-and-forward-at-relay}.
\subsubsection{Problem Formulation}\label{secIV_subsecA_subsubsec1}

Consider the symmetric-rate $R^{\text{CoF}}_{\text{sym}}$ as given in Proposition~\ref{proposition-achievable-sum-rate-compute-and-forward-at-relay}. The optimization problem can be stated as: 
\begin{subequations}\label{statement-optimization-problem-first-strategy}
\begin{eqnarray}
\text{(OP)\::} && \max_{\{\dv a_m\}_{m=1}^M,\:\boldsymbol{\beta}}\:\: \min \left\{\min_{m}\:R(\dv a_m,\dv h_m,\boldsymbol{\beta}),\: R_o \right\}\\
&& \quad \textrm{s. t.} \qquad -\sqrt{\frac{P_{u_m}}{P}} \leq \beta_{u_m} \leq \sqrt{\frac{P_{u_m}}{P}}\\
&& \quad \qquad \qquad \text{rank}(\dv A)=M.
\end{eqnarray}
\end{subequations}
The optimization problem (OP) is a non-linear mixed integer optimization problem. Thus it is hard to find $\boldsymbol{\beta}$ and $\{\dv a_m\}_{m=1}^M$ jointly in a reasonable time \cite{SZV14}. Therefore, we propose an iterative optimization where we find appropriate precoding vector $\mathrm{\boldsymbol{\beta}}$ and integer coefficients $\{\dv a_m\}_{m=1}^M$ alternately. Let us denote by $R^{\text{CoF}}_{\text{sym}}[\iota]$ the value of the symmetric-rate at some iteration $\iota \geq 0$. We develop ``Algorithm OP" to find the appropriate $\mathrm{\boldsymbol{\beta}}$ and $\{\dv a_m\}_{m=1}^M$ that maximize $R^{\text{CoF}}_{\text{sym}}$. 

As described in ``Algorithm OP", we find the appropriate $\boldsymbol{\beta}$ and $\dv A$, alternately. 
The iterative process in ``Algorithm OP" terminates if either one of the following conditions holds: i) $\|\mathrm{\boldsymbol{\beta}}^{(\iota)}-\mathrm{\boldsymbol{\beta}}^{(\iota-1)}\|$ and $|R^{\text{CoF}}_{\text{sym}}[\iota]-R^{\text{CoF}}_{\text{sym}}[\iota-1]|$ are smaller than prescribed small strictly positive constants $\epsilon_1$ and $\epsilon_2$, respectively --- in this case, the optimized value of the symmetric-rate is $R^{\text{CoF}}_{\text{sym}}[\iota]$ and is obtained using $\boldsymbol{\beta}^{\star}=\mathrm{\boldsymbol{\beta}}^{(\iota)}$ and $\dv A^{\star}= \dv A^{(\iota)}$ ii) $\text{rank}(\dv A^{(\iota)})< M$--- in this case,  if $\iota=1$  the optimized value of the symmetric-rate is obtained using $\boldsymbol{\beta}^{\star}=\dv 0$ and $\dv A^{\star}= \dv 0$ otherwise $\boldsymbol{\beta}^{\star}=\mathrm{\boldsymbol{\beta}}^{(\iota-1)}$ and $\dv A^{\star}= \dv A^{(\iota-1)}$.

\begin{algorithm}[h!] 
\renewcommand{\thealgorithm}{OP}
{\fontsize{8}{8}\selectfont
\caption{\small{Iterative algorithm to compute $R^{\text{CoF}}_{\text{sym}}$ as given by (\ref{achievable-sum-rate-compute-and-forward-at-relay})}}\label{al:1}
\begin{algorithmic}[1]
\State Initialization: set $\iota =1$ and $\mathrm{\boldsymbol{\beta}}=\mathrm{\boldsymbol{\beta}}^{(0)}$, where $\boldsymbol{\beta}^{(0)}$ is a given initial value
\State Set $\mathrm{\boldsymbol{\beta}}=\mathrm{\boldsymbol{\beta}}^{(\iota-1)}$ in \eqref{statement-optimization-problem-first-strategy}, and solve the obtained problem as described in Section \ref{secIV_subsecA_subsubsec2}. Denote by $\dv A^{(\iota)}$ the found  $\dv A$
\State \textbf{If} $\text{rank}(\dv A^{(\iota)}) = M$ 
\State \quad Set $\dv A=\dv A^{(\iota)}$ in \eqref{statement-optimization-problem-first-strategy}, and solve the obtained problem using Algorithm OP-1 given below. Denote by $\boldsymbol{\beta}^{(\iota)}$ the found $\boldsymbol{\beta}$
\State \quad Increment the iteration index as $\iota=\iota+1$, and go back to Step 2
\State \quad Terminate if $\|\mathrm{\boldsymbol{\beta}}^{(\iota)}-\mathrm{\boldsymbol{\beta}}^{(\iota-1)}\|\leq \epsilon_1$, $|R^{\text{CoF}}_{\text{sym}}[\iota]-R^{\text{CoF}}_{\text{sym}}[\iota-1]|\leq \epsilon_2$
\State \textbf{Else}
\State \quad \textbf{If} $\iota =1$
\State \quad \quad Terminate and set $\boldsymbol{\beta} = \dv 0$ and $\dv A=\dv 0$
\State \quad \textbf{Else}
\State \quad \quad Terminate and set $\boldsymbol{\beta}$ = $\boldsymbol{\beta}^{(\iota-1)}$ and $\dv A=\dv A^{(\iota-1)}$
\State \quad \textbf{End}
\State \textbf{End}
\end{algorithmic}
}
\end{algorithm}
We should note that by considering different initial values $\boldsymbol{\beta}^{(0)}$ a higher transmission rate can be obtained and the probability to get full rank linear combinations can be increased.

\subsubsection{Integer Coefficients Optimization}\label{secIV_subsecA_subsubsec2}

In this section, we search for the integer coefficients $\{\dv a_m\}_{m=1}^M$ for a given $\boldsymbol{\beta}$. The optimization problem (OP) can be equivalently written as 
\begin{subequations}\label{optimizing-integer-coefficients-for-given-beta}
\begin{eqnarray}
\min_{\{\dv a_m\}_{m=1}^M,\: \Delta_1}&&\Delta_1\\
\textrm{s. t.}&& \Delta_1 \geq \dv a_m^T\Omega_m \dv a_m\\  
&& \Delta_1 \geq 2^{-R_o}
\end{eqnarray}
\end{subequations}
where $\Delta_1 \in \mathbb{R}$ is simultaneously a slack variable and the objective function, and $\Omega_m = \dv I_M  -\frac{P(\boldsymbol{\beta} \circ \dv h_m)(\boldsymbol{\beta} \circ \dv h_m)^T}{N+P\|\boldsymbol{\beta} \circ \dv h_m\|^2} \in \mathbb{R}^{M \times M}$.

The optimization problem \eqref{optimizing-integer-coefficients-for-given-beta} is a mixed integer quadratic programming (MIQP) problem \cite{SZV14} \cite{F95} and can easily and efficiently be solved using branch and bound method \cite{W98}.

\subsubsection{Precoding Allocation}\label{secIV_subsecA_subsubsec3}

In this section, we optimize the precoding vector $\boldsymbol{\beta}$ for given $\{\dv a_m\}_{m=1}^M$. Again, we can rewrite the optimization problem (OP), for $m = 1,\dots,M$, as
\begin{subequations}\label{optimizing-betas-for-given-integer-coefficients}
\begin{eqnarray}
\min_{\mathrm{\boldsymbol{\beta}},\:\Delta_2}&& \Delta_2 \\
\textrm{s. t.}&& \Delta_2 \geq \|\dv a_m\|^2-\frac{P((\boldsymbol{\beta} \circ \dv h_m)^T \dv a_m)^2}{N+P\|\boldsymbol{\beta} \circ \dv h_m\|^2} \label{power_allocation_1}\\
&& \Delta_2 \geq 2^{-R_o} \label{power_allocation_3}\\
&& -\sqrt{\frac{P_{u_m}}{P}} \leq \beta_{u_m} \leq \sqrt{\frac{P_{u_m}}{P}} \label{power_allocation_4}
\end{eqnarray}
\end{subequations}
where $\Delta_2\in \mathbb{R}$ is simultaneously a slack variable and the objective function.
\noindent The optimization problem in \eqref{optimizing-betas-for-given-integer-coefficients} is non-linear and non-convex. This problem can be formulated as a complementary geometric program (CGP) \cite{SZV14} \cite{CTPOJ07}  and can be solved easily and efficiently as described in \cite{SZV14}. To solve a CGP problem, we need to transform it into a geometric program (GP). This means that the variables in the optimization problem should be all positive, and the objective function and the constraints should be posynomials. We define $\dv c=[c_{u_1},\ldots c_{u_M}]^T \in \mathbb{R}^M$ and $\boldsymbol{\delta}=[\delta_{u_1},\ldots, \delta_{u_M}]^T \in \mathbb{R}^M$, such that  $c_{u_m} > \sqrt{P_{u_m}/P}$ and $\delta_{u_m}=\beta_{u_m}+c_{u_m}$ for $m=1,\ldots, M$. It can easily be seen that the elements of $\boldsymbol{\delta}$ are all strictly positive. Hence, the optimization problem \eqref{optimizing-betas-for-given-integer-coefficients} can be written in the following form,
\begin{subequations}\label{constraints-equivalent-form-optimizing-betas-for-given-integer-coefficients}
\begin{eqnarray}
\min_{\mathrm{\boldsymbol{\delta}},\: \Delta_2}&&\Delta_2 \\
\textrm{s. t.}&&\frac{f_m(\mathrm{\boldsymbol{\delta}},\Delta_2)}{g_m(\mathrm{\boldsymbol{\delta}},\Delta_2)}\leq 1 \label{CGP_beta}\\
&& \frac{2^{-R_o}}{\Delta_2}\leq 1\\
&& -\sqrt{\frac{P_{u_m}}{P}}+ c_{u_m} \leq \delta_{u_m} \leq \sqrt{\frac{P_{u_m}}{P}}+ c_{u_m},
\end{eqnarray}
\end{subequations}
where the constraints in \eqref{CGP_beta} correspond to the constraints in \eqref{power_allocation_1}, and $f_m(\mathrm{\boldsymbol{\delta}},\Delta_2)$ and $g_m(\mathrm{\boldsymbol{\delta}},\Delta_2)$ are posynomial functions. It is easy to see that the constraints in \eqref{CGP_beta} are not posynomial since a ratio of posynomial functions is not posynomial \cite{CTPOJ07}. Therefore, we use Lemma~1 of \cite{SZV14} to approximate the functions $g_m(\mathrm{\boldsymbol{\delta}},\Delta_2)$ with monomials $\tilde{g}_m(\mathrm{\boldsymbol{\delta}},\Delta_2)$ around some initial value. We should note that the ratio between posynomial and monomial can be upper bounded by a posynomial \cite{CTPOJ07}. Thus, the optimization problem \eqref{constraints-equivalent-form-optimizing-betas-for-given-integer-coefficients} is now a GP problem and can be solved easily using an interior point approach. To improve the accuracy of the approximation, the found solution of the GP problem is used as an initial value to transform again the CGP into a new GP problem. This process is repeated until convergence to a stationary point. 
The problem of finding $\boldsymbol{\delta}$ for given $\{\dv a_m\}_{m=1}^M$ is described in ``Algorithm OP-1".
\begin{algorithm}[h!]
\renewcommand{\thealgorithm}{OP-1}
{\fontsize{8}{8}\selectfont
\caption{\small{Precoding allocation for $R^{\text{CoF}}_{\text{sym}}$ as given by (\ref{achievable-sum-rate-compute-and-forward-at-relay})}}
\begin{algorithmic}[1]
\State Set $\mathrm{\boldsymbol{\delta}}^{(0)}$ to some initial value. Compute $\Delta_2^{(0)}$ using $\mathrm{\boldsymbol{\delta}}^{(0)}$ and set $\iota_2 =1$
\State Approximate $g(\mathrm{\boldsymbol{\delta}}^{(\iota_2)},\Delta_2^{(\iota_2)})$ with $\tilde{g}(\mathrm{\boldsymbol{\delta}}^{(\iota_2)},\Delta_2^{(\iota_2)})$ around $\mathrm{\boldsymbol{\delta}}^{(\iota_2-1)}$ and $\Delta_2^{(\iota_2-1)}$ using Lemma~1 of \cite{SZV14} 
\State Solve the resulting approximated GP problem using an interior point approach. Denote the found solutions as $\mathrm{\boldsymbol{\delta}}^{(\iota_2)}$ and $\Delta_2^{(\iota_2)}$
\State Increment the iteration index as $\iota_2=\iota_2+1$ and go back to Step 2 using $\mathrm{\boldsymbol{\delta}}$ and $\Delta_2$ of step 3
\State Terminate if $\|\mathrm{\boldsymbol{\delta}}^{(\iota_2)}-\mathrm{\boldsymbol{\delta}}^{(\iota_2-1)}\|\leq \epsilon_1$
\end{algorithmic}
}
\end{algorithm}

\section{Numerical Examples}\label{secV}

In this section, we provide some numerical examples where we measure the performance of the coding strategy using symmetric outage rate. We compare our coding strategy with the traditional strategies for $M=2$. Also, we consider different algorithms and we compare them with the proposed algorithm ``Algorithm OP". The symmetric outage rate is given by \cite{NG11a},

\begin{equation}
R^{\text{CoF}}_{\text{Out}} = \text{sup} \{R:\rho_{\text{Out}}(R) \leq \rho \}
\end{equation}
where $\rho_{\text{Out}}(R)$ is the outage probability and is given by
\begin{equation}
\rho_{\text{Out}}(R) = \text{Pr}(R^{\text{CoF}}_{\text{sym}}<R).
\end{equation}

Throughout this section, we assume that the channel coefficients are modeled with independent and randomly generated variables, each generated according to a zero-mean Gaussian distribution with variance $\sigma^2_{u_i,r_j}$, for $i,j=1,2$. Also, we set $P_{u_1} = 20$ dBW, $P_{u_2} = 20$ dBW, $P = 20$ dBW, and $\rho = 1/4$. 

\begin{figure*}
        \centering
        \begin{subfigure}[b]{0.35\textwidth}
                \includegraphics[width=\textwidth]{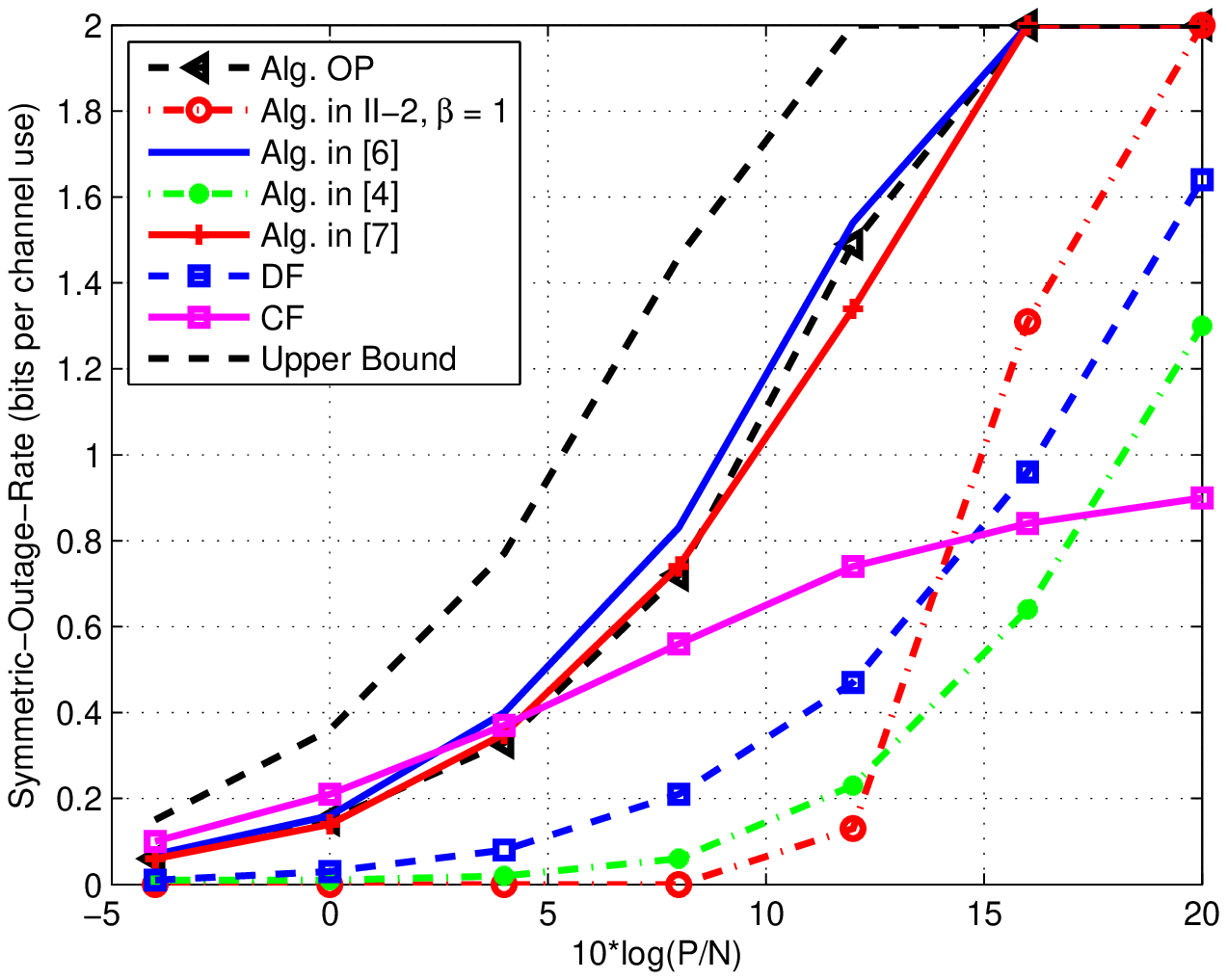}
                \caption{}
                \label{fig:Out1}
        \end{subfigure}%
        ~ \hspace{-0.75 cm}
        \begin{subfigure}[b]{0.35\textwidth}
                \includegraphics[width=\textwidth]{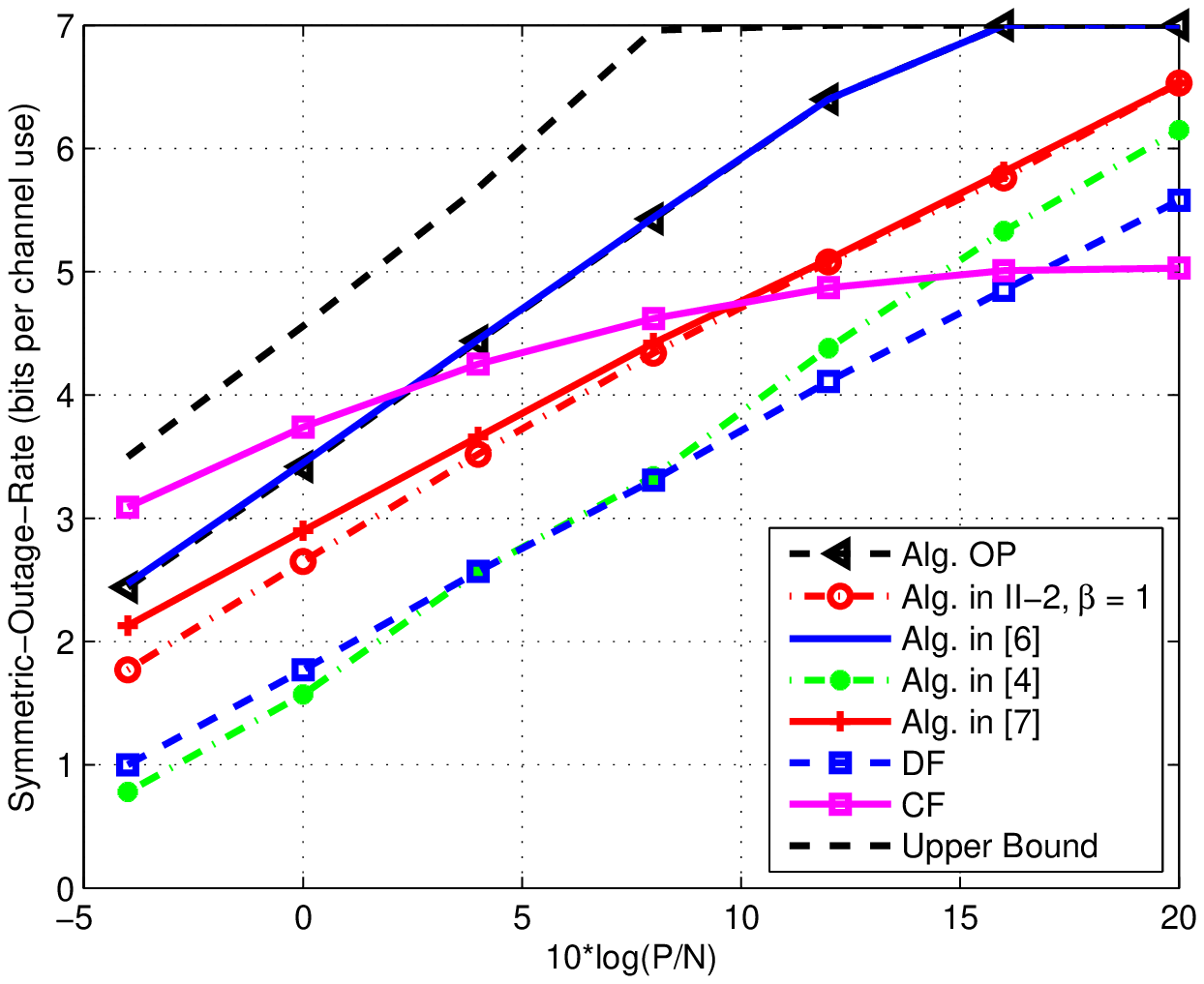}
                \caption{}
                \label{fig:Out10}
        \end{subfigure}
        ~ \hspace{-0.75 cm}
        \begin{subfigure}[b]{0.35\textwidth}
                \includegraphics[width=\textwidth]{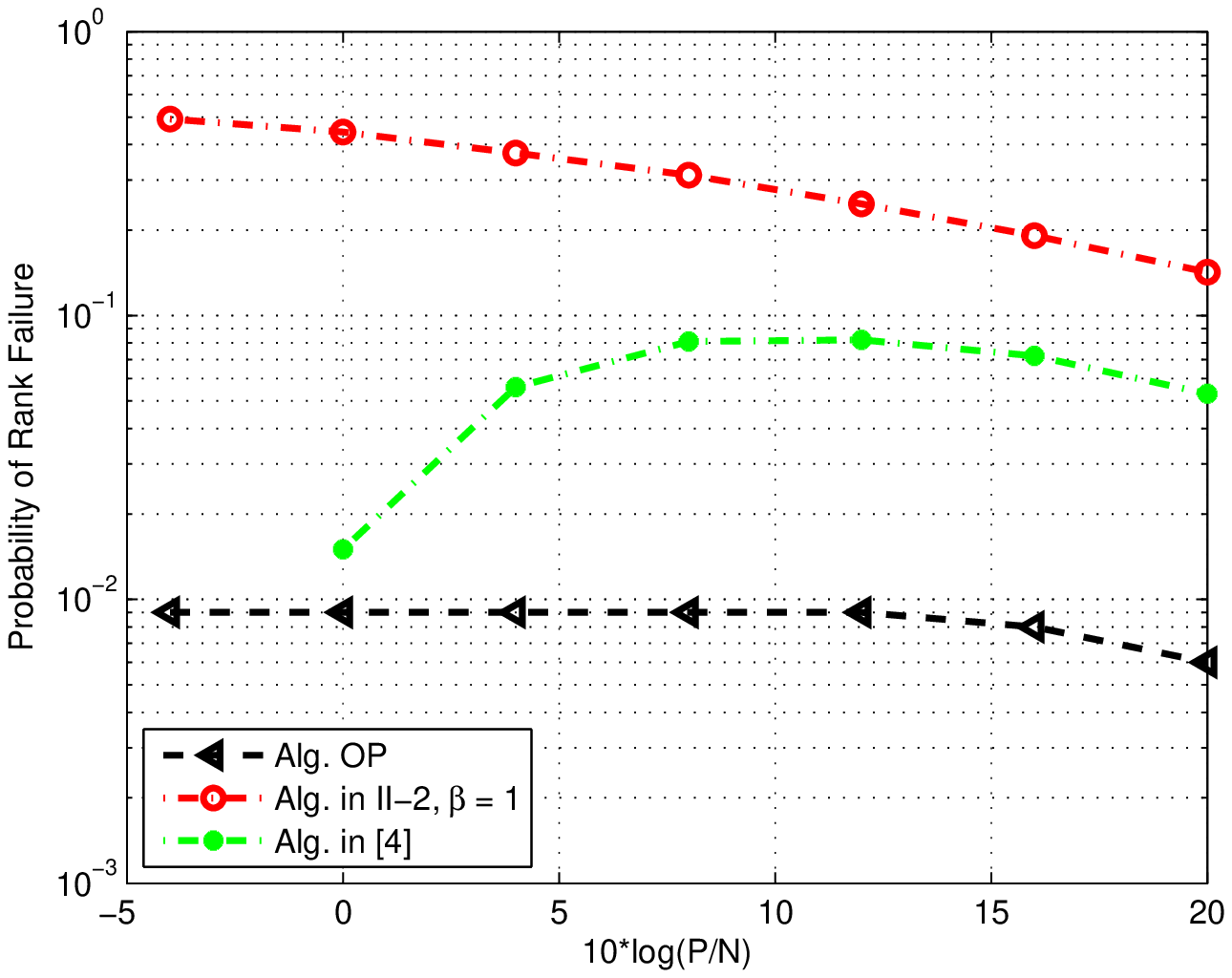}
                \caption{}
                \label{fig:Prob}
        \end{subfigure}
        \caption{\small{(a) Symmetric outage rates, $R_o = 2$ bits per channel use, $\sigma^2_{u_1,r_1}= \sigma^2_{u_2,r_1}=\sigma^2_{u_1,r_2}=\sigma^2_{u_2,r_2}=0$ dB. (b) Symmetric outage rates, $R_o = 7$ bits per channel use, $\sigma^2_{u_1,r_1}= \sigma^2_{u_2,r_1}=\sigma^2_{u_1,r_2}=\sigma^2_{u_2,r_2}=20$ dB. (c) Probability of rank failure, $R_o = 2$ bits per channel use, $\sigma^2_{u_1,r_1}= \sigma^2_{u_2,r_1}=\sigma^2_{u_1,r_2}=\sigma^2_{u_2,r_2}=0$ dB.}}\label{fig:sym-outage-rate}
\end{figure*}

Figures~\ref{fig:Out1} and \ref{fig:Out10} show the symmetric outage rate obtained using the CoF approach under different optimization algorithms: i) using ``Algorithm OP", i.e., $R^{\text{CoF}}_{\text{Out}}$ (Alg. OP), ii) using the integer coefficients algorithm described in \ref{secIV_subsecA_subsubsec2} with $\boldsymbol{\beta}=\dv 1$, i.e., $R^{\text{CoF}}_{\text{Out}}$ (Alg. in \ref{secIV_subsecA_subsubsec2}, $\boldsymbol{\beta}=\dv 1$), iii) using the algorithm described in \cite{SZV14}, i.e., $R^{\text{CoF}}_{\text{Out}}$ (Alg. in \cite{SZV14}), iv) using the integer coefficients algorithm described in \ref{secIV_subsecA_subsubsec2} but forcing relay $m$ to compute an equation with $a_{mm}\neq 0$, i.e., $R^{\text{CoF}}_{\text{Out}}$ (Alg. in \cite{NG11a}) v) using the algorithm described in \cite{WC12} i.e., $R^{\text{CoF}}_{\text{Out}}$ (Alg. in \cite{WC12}) as functions of $10\log(P/N)$. The figures also show the symmetric outage rates obtained using DF, CF and the upper bound as given in \cite{NG11a}.

For the example shown in Figure~\ref{fig:Out1}, we observe that ``Algorithm OP" achieves a symmetric outage rate slightly less than what is obtained using the algorithm in \cite{SZV14}. Also, we observe that ``Algorithm OP" has a performance similar to that in \cite{WC12} at low values of $P/N$, however it has higher performance at mid and high values of $P/N$. Recall that, in \cite{SZV14} and \cite{WC12}, the integer coefficient vectors are jointly computed among the relays. In these methods, each relay finds a candidate set that contains several integer vectors. From those candidate sets, the relays jointly select from each set an integer vector to construct a full rank matrix that maximizes the transmission rate. In order to jointly select the integer coefficient vectors, the relays need either to signal their candidate sets to each others or to transmit them to a central controller. This makes those methods not practical for a large number of users and relays and makes the proposed method more practical and efficient. Moreover, the complexity to select the independent integer vectors from the candidate sets is $\mathcal{O}(T^M)$ where $T$ is the number of integer vectors in each set. In contrast, in the proposed method, it is zero since each relay independently computes an integer vector. Also, we observe that ``Algorithm OP" outperforms the other described algorithms and that CoF strategy, in this regime, has better performance than standard DF and CF as it has been shown in \cite{NG11a}.

Figure~\ref{fig:Out10} depicts the same curves for other channel variances. In this case, we observe that both algorithms ``Algorithm OP" and the one in \cite{SZV14} have the same performance. Also, we observe that ``Algorithm OP" significantly outperforms the algorithm in \cite{WC12}.    

In Figure~\ref{fig:Prob}, we observe that the probability that the linear combinations are not full rank is quite small using ``Algorithm OP" compared with the other algorithms. Hence, we notice that precoding allocation can help to increase the transmission rate, to decrease the probability of rank failure, and to reduce the complexity at the relays.




\section{Conclusion}\label{secVI}
In this paper, we consider a system where multiple users communicate with a destination with the help of multiple half-duplex relays. The relays use the CoF strategy where each relay decodes an integer-valued linear combination that relates the transmitted codewords and then forwards it to the destination. Given these linear combinations, the destination may or may not recover the transmitted messages since the linear combinations are not always full rank. To reduce the probability of rank failure at the destination and to maximize the transmission rate, we consider precoding allocation at the users. The analysis shows the advantage of the precoding technique over other techniques and the advantage of CoF strategy over the traditional strategies. 

%



\bibliographystyle{IEEEtran}
\bibliography{bibfile}

\begin{thebibliography}{10}
\providecommand{\url}[1]{#1}
\csname url@samestyle\endcsname
\providecommand{\newblock}{\relax}
\providecommand{\bibinfo}[2]{#2}
\providecommand{\BIBentrySTDinterwordspacing}{\spaceskip=0pt\relax}
\providecommand{\BIBentryALTinterwordstretchfactor}{4}
\providecommand{\BIBentryALTinterwordspacing}{\spaceskip=\fontdimen2\font plus
\BIBentryALTinterwordstretchfactor\fontdimen3\font minus
  \fontdimen4\font\relax}
\providecommand{\BIBforeignlanguage}[2]{{%
\expandafter\ifx\csname l@#1\endcsname\relax
\typeout{** WARNING: IEEEtran.bst: No hyphenation pattern has been}%
\typeout{** loaded for the language `#1'. Using the pattern for}%
\typeout{** the default language instead.}%
\else
\language=\csname l@#1\endcsname
\fi
#2}}
\providecommand{\BIBdecl}{\relax}
\BIBdecl

\bibitem{ACLY00}
R.~Ahlswede, N.~Cai, S.-Y.~R. Li, and R.~W. Yeung, ``Network information
  flow,'' \emph{{IEEE} Trans. Inf. Theory}, vol.~46, pp. 1204--1216, 2000.

\bibitem{FF56}
L.~R. Ford and D.~R. Fulkerson, ``Maximal flow through a network,'' \emph{Can.
  J. Math.}, vol.~8, pp. 399--404, 1956.

\bibitem{FS07}
C.~Fragouli and E.~Soljanin, \emph{Network coding fundamentals}.\hskip 1em plus
  0.5em minus 0.4em\relax Found. Trends Commun. Info. Theory, 2007.

\bibitem{NG11a}
B.~Nazer and M.~Gastpar, ``Compute-and-forward: harnessing interference through
  structured codes,'' \emph{{IEEE} Trans. Inf. Theory}, vol.~57, pp.
  6463--6486, 2011.

\bibitem{KMT08}
S.~J. Kim, P.~Mitran, and V.~Tarokh, ``Performance bounds for bi-directional
  coded cooperation protocols,'' \emph{{IEEE} Trans. Inf. Theory}, vol. IT-54,
  pp. 5253--5241, Nov. 2008.

\bibitem{SZV14}
M.~El-Soussi, A.~Zaidi, and L.~Vandendorpe, ``Compute-and-forward on a
  multiaccess relay channel: Coding and symmetric-rate optimization,''
  \emph{{IEEE} Trans. Wireless Communications}, vol.~13, pp. 1932--1947, 2014.

\bibitem{WC12}
L.~Wei and W.~Chen, ``Compute-and-forward network coding design over
  multi-source multi-relay channels,'' \emph{{IEEE} Trans. Wireless
  Communications}, vol.~11, pp. 3348--3357, 2012.

\bibitem{CFB14}
\BIBentryALTinterwordspacing
Z.~Chen, P.~Fan, and K.~B. Letaief, ``Compute-and-forward: Optimization over
  multi-source-multi-relay networks,'' 2014. [Online]. Available:
  \url{http://arxiv.org/pdf/1406.1081.pdf}
\BIBentrySTDinterwordspacing

\bibitem{F95}
C.~A. Floudas, \emph{Nonlinear and mixed integer optimization}.\hskip 1em plus
  0.5em minus 0.4em\relax Oxford University Press, 1995.

\bibitem{W98}
L.~A. Wolsey, \emph{Integer programming}.\hskip 1em plus 0.5em minus
  0.4em\relax John Wiley and Sons, 1998.

\bibitem{CTPOJ07}
M.~Chiang, C.~W. Tan, D.~P. Palomar, D.~O'Neill, and D.~Julian, ``Power control
  by geometric programming,'' \emph{{IEEE} Trans. Wireless Communications},
  vol.~6, pp. 2640--2651, July 2007.

\end{thebibliography}

\end{document}